\documentclass[a4paper,12pt]{article}

\usepackage[T2A]{fontenc}
\usepackage{amsmath,amssymb}
\usepackage{cite}
\usepackage[linktocpage=true,plainpages=false,pdfpagelabels=false]{hyperref}

\usepackage[utf8]{inputenc}
\usepackage[english]{babel}

\usepackage{xfrac}
\usepackage{subfig}
\usepackage{array}
\usepackage{tabularx} 

\usepackage[svgnames,table]{xcolor}

\definecolor{light-gray}{gray}{0.95}

\newcommand{\de}{\delta}

\newcommand{\gm}{\gamma}
\newcommand{\al}{\alpha}

\newcommand{\vp}{\varphi}
\newcommand{\la}{\lambda}

\newcommand{\om}{\omega}
\newcommand{\tta}{\theta}
\newcommand{\bt}{\beta}
\newcommand{\sh}{\sinh}
\newcommand{\ch}{\cosh}

\begin{document}

\title{Classification of ten-dimensional embeddings of spherically symmetric static metrics}
\author{S.~S.~Kuptsov\thanks{E-mail: skup.sov@yandex.ru},
S.~A.~Paston\thanks{E-mail: pastonsergey@gmail.com},
A.~A.~Sheykin\thanks{E-mail: a.sheykin@spbu.ru}\\
{\it PDMI RAS, Saint Petersburg, Russia}\footnotemark[1]\\
{\it Saint Petersburg State University, Saint Petersburg, Russia}\footnotemark[2] \footnotemark[3]
}
\date{}
\maketitle

\begin{abstract}
The group-theoretic method for constructing symmetric isometric embeddings is used to describe all possible four-dimensional surfaces in flat $(1,9)$-dimensional space, whose induced metric is static and spherically symmetric. For such surfaces, we propose a classification related to the dimension of the elementary blocks forming the embedding function. All suitable 52 classes of embeddings are summarized in one table and analyzed for the unfolding property (wich means that the surface does not belong locally to some subspace of the ambient space), as well as for the presence of smooth embeddings of the Minkowski metric. The obtained results are useful for the analysis of the equations of motion in the Regge-Teitelboim embedding gravity, where the presence of unfolded embeddings of the Minkowski metric is essential.
\end{abstract}

\newpage

\section{Introduction}\label{}
As we know from the Janet-Cartan-Friedman theorem \cite{fridman61}, any  $d$-dimensional Riemannian space can be locally isometrically embedded in any other Riemannian space of dimension greater than or equal to $n=d\left( d+1 \right)/2$. This fact opens up the possibility that any (pseudo-)Riemannian space can be viewed as a surface in an ordinary (pseudo-)Euclidean space of suitable dimension and signature. In this case, the metric of the embedded space coincides with the one that is induced by the metric of the ambient space on the corresponding surface. If $y^a\left( x^\mu \right)$ is an embedding function describing such a surface, then the formula for the induced metric is:

\begin{equation}
	g_{\mu\nu} = \partial_\mu y^a \partial_\nu y^b\, \eta_{ab}, \label{g}
\end{equation}
where $\mu,\nu = 0,\ldots,d-1$; $a,b = 0,\ldots,n-1$; $\eta_{ab}$ -- metric of the (pseudo-)Euclidean space.

Looking from this perspective not only gives more clarity to geometry of a particular Riemannian space (e.g. the Fronsdal embedding \cite{frons} of Schwarzschild black hole helps to understand its topology) but also simplifies some calculations. In addition to various applications to GR, such as the determination of the quasi-local mass \cite{0805.1370} and the calculation of the Hawking temperature of spaces with the horizon \cite{deserlev99, statja36}, embeddings can also be used to modify GR. The idea of our spacetime as a surface in a ten-dimensional flat space determines one of these modifications -- the Regge-Teitelboim embedding theory \cite{regge}. This theory itself is an extension of GR \cite{deser,pavsic85,davkar}, and when additional constraints are imposed, it can be understood as an equivalent reformulation of GR (this is the view used in the pioneering work \cite{regge} in the hope of simplifying the procedure for quantization of gravity). The independent variable in the embedding theory approach is not the metric, but the embedding function $y^a$, while the usual Einstein-Hilbert expression is taken as an action, in which the formula for the induced metric \eqref{g} is substituted. Thus, the embedding theory arises from GR via differential transformations of field variables \cite{statja60}, as in the case of mimetic gravity \cite{mukhanov, mimetic-review17}, which has been frequently discussed in the recent years. At the same time, unlike its analog in mimetic gravity, the change of variables in embedding theory has a clear geometric meaning inspired by string theory \cite{Sheykin_2020}.

The Janet-Cartan-Friedman theorem guarantees the existence of the local embedding but does not give any clues about its explicit form. The search for an embedding $y^a$ that corresponds to a given Riemannian space, i.e. is a solution to the equation \eqref{g} with a given metric, has to be done case-by-case. However, there are a few general methods for solving such problems that are applicable in some situations. For example, in the case of small codimension, the analysis of the second fundamental form of the surface works well, in which the formula \eqref{g} is replaced by a system of Gauss-Codazzi equations (see \cite{schmutzer}, section 37). We will be interested in the case of codimension 6, where this method is not so good due to the large number of unknown functions. Instead, we employ another method proposed in \cite{statja27} for construction all possible embeddings possessing any given symmetry. The same symmetry will automatically be possessed by the metric induced on such an embedding (on the right side of \eqref{g}). Having a complete set of such embeddings, we can hope that the variables in the equation \eqref{g} will separate, and at least for some symmetric embeddings it will be possible to solve it explicitly. At the same time, it should be understood that with this approach a part of the embeddings that satisfy \eqref{g} could not be obtained, because a surface that is asymmetric from the point of view of external geometry may nevertheless have a symmetric induced metric. For example, isometric curved local embeddings of a sphere have axial rather than spherical symmetry (see \cite{rashevsky_1950}, section 97). Nevertheless, there is a modification of the mentioned method, which allows one to work with this kind of embeddings as well \cite{statja70}.

We will be interested in embeddings of four-dimensional spacetime that are spherically symmetric and invariant under translations in time, i.e. having symmetry under transformations from the group $SO(3)\times T^1$. It is worth noting that although the Janet-Cartan-Friedman theorem guarantees the existence of embeddings of a given metric into ambient spaces of dimension 10 or more, the existence of symmetries allows one to lower this minimal number of dimensions. In particular, it can be shown that the minimal dimension of the ambient space allowing $SO(3)\times T^1$-symmetric four-dimensional embeddings is reduced to six, and in some cases to five (see \cite{schmutzer}, section 37.3). The corresponding embeddings in a six-dimensional space satisfying the Einstein equation are listed in \cite{kasner3, frons, fujitani, davidson, statja27}. 

The scope of this paper is to obtain a complete classification of $SO(3)\times T^1$-symmetric embeddings into full $n=10$, while not limiting ourselves to solutions of the Einstein equations. Such a classification is useful, in particular, for a perturbative analysis of Regge-Teitelboim equations if the background distribution of matter is considered to be spherically symmetric and static in some approximation, i.e. having $SO(3)\times T^1$ symmetry. A special case of such embedding has already been used as a background in the study of linearized equations of motion of the embedding theory \cite{Kuptsov_2022}. Other embeddings given by this classification can also be used as a background both in the analysis of linearized equations and in the study of the non-relativistic limit of the embedding theory \cite{statja68}, where the problem of choosing a background embedding also arises. For these problems, the embedding has to be "present"\ in all ten directions, i.e. it must be impossible to present any subspace (in particular, a six-dimensional subspace) that would completely include the surface under study. Mathematically, this "presence"\ is reduced to the verification of the unfolding property of the embedding \cite{Zaitseva_2021}, see Section \ref{rabota} for details.

In the Section \ref{infa 2012}, we will describe the method of constructing symmetric embeddings proposed in \cite{statja27}, discuss its application to our task, and describe in detail the results of applying the method for the group $SO(3)\times T^1$. In Section \ref{rabota} we will use these results to classify the $SO(3) \times T^1 $ symmetric ten-dimensional embeddings of interest to us, which we will call symmetric embeddings of the second type, and also discuss their properties: unfolding and the possible existence of everywhere smooth embeddings of the Minkowski metric.

\section{Method for constructing symmetric embeddings} \label{infa 2012}
Let us first define the symmetry of the surface described by the embedding function $y^a(x^\mu)$. We will assume that the surface is symmetric with respect to the group $G$ if in the symmetry group of an ambient space $SO(n_+,n_-) \ltimes T^n$ (here $n=n_++n_-$) there is a subgroup isomorphic to $G$ under the action of which the surface transforms into itself. This definition has several important consequences. First, the choice of an isomorphic subgroup uniquely determines how $G$ acts on surface points: any group transformation must be a concrete combination of rotation and translation of the vector $y^a$ in the ambient space. Surfaces symmetric under the action of one isomorphic subgroup are not symmetric under the action of any other. It is important to keep this in mind, in particular, when considering $SO(3)$-symmetric surfaces, when the action of the group may turn out to be nontrivial by construction, i.e. not reduced to the transformation of only the standard angular variables $\tta$ and $\vp$ at fixed others (in particular -- at fixed radial coordinate $r$). An example of this will be given below. Second, this action of the group is an isometry with respect to the induced metric \eqref{g}, and in this sense $G$ is automatically also a symmetry group of the metric. Third, it follows from the definition that symmetric surfaces can be reconstructed to some extent from one given point of it: acting on it by group transformations, one can obtain a manifold that either coincides with the whole surface or is some submanifold of it, i.e. the orbit of the group of transformations. This idea forms a basis of the method proposed in \cite{statja27} for constructing embedding functions $y^a(x^\mu)$ that correspond to surfaces of a given symmetry. We will describe this method, omitting the technical details and focusing only on the results we need.

Before doing so, it is important to distinguish between two types of symmetric surfaces, which helps us to pose the problem correctly. Surfaces of the first types are ones any two points of those can be converted into each other by a symmetry transformation. Symmetric surfaces of the first type coincide with some orbit of the subgroup of $SO(n_+,n_-)\ltimes T^n$ isomorphic to $G$. Surfaces of the second type are split into a set of orbits, that is, into a system of their own submanifolds, each of which in itself is a symmetric surface. The dimension of such symmetric submanifolds will be further denoted by the letter $\xi$. The corresponding part of the degrees of freedom of the symmetric surface is associated with the group parameters, and the other part does not depend on the group in any way and only numbers the orbits. The details of the application of the mentioned method may vary depending on what type of symmetric surfaces we want to obtain. Below we describe in detail one nontrivial example of a surface of the first type. But it is more important for us to obtain a classification of surfaces of the second type, the justification of which will be given a little later.

Let us move on to a brief description of the method by which we will construct such a classification. The symmetry group $G$ of dimension $m$, the dimension of the ambient space $n$, the dimension of the surface $d$ and the dimension of its symmetric submanifolds $\xi$ (for surfaces of the first type $\xi=d$, for the second type $\xi<d$) we set as input data for the method. The method essentially consists of counting and constructing of all possible symmetric surfaces that correspond to these input data. The algorithm is the following. First of all, we choose a representation of the group $G$ of dimension $n$ whose matrices belong also to the fundamental representation $SO(n_+,n_-)\ltimes T^n$. Then we select an initial vector in the space of this representation (it also plays the role of the ambient space). We act on it by representation matrices that correspond to all elements of $G$, thus obtaining a symmetric surface of the first type. Then the method starts to vary depending on the relations between $m$, $\xi$ and $d$.

Suppose first that the initial vector has been chosen in a general case. This means that the variation of each group parameter leads to the displacement of this vector, and the dimension of the obtained surface coincides with $m$. If we are interested in surfaces of the first type with $m=\xi=d$, then a suitable surface has been found, and the procedure can now be repeated for other initial vectors and representations. If we are interested in the case $m=\xi<d$, then the desired surface must be of the second type. Then we interpret the surface obtained by acting on the initial vector as a symmetric submanifold of the desired surface. The missing $d-\xi$ degrees of freedom which are invariant under the action of the group can be included by assigning an arbitrary dependence on the corresponding coordinates to the initial vector itself.

All cases $m<\xi$ have no solutions -- it is impossible to construct an orbit of a group of dimension greater than the dimension of the group itself. However, it is possible that surfaces of both types exist when $m>\xi$. To describe them, we need to learn how to construct surfaces of the first type of dimension smaller than the dimension of the group. In other words, we need to eliminate a part of the degrees of freedom of the surface associated with $G$. This can be done if the representation of the group allows such a choice of the initial vector that corresponds to some stabilizer subgroup of dimension $m-\xi$. In this case, the initial vector remains stationary under the action of the representation matrices corresponding to the variation of $m-\xi$ group parameters, and the action of all the remaining matrices on it forms a surface of the first type of dimension $\xi<m$. Surfaces of the second type, as before, can be obtained by making the initial vector dependent on coordinates untouched by the symmetry.

Since any symmetric surface must be obtained in one of the described ways from some initial vector, the problem of classifying them actually reduces to the problem of classifying suitable representations of $G$ and initial vectors. For all physically interesting groups, the theory of representations is well developed, and complete classifications of representations are known. We only need to select among them those that realize transformations from $SO(n_+,n_-)\ltimes T^n$, and in addition to this allow vectors with stationary subgroup in the case $\xi<m$. 

\subsection{The case of static spherically symmetric metrics}
Let us now discuss our specific formulation of the problem and the corresponding input data for the method. We are interested not so much in the classification of symmetric surfaces as in the classification of symmetric embeddings of given metrics, when their very form dictates the action of the symmetry group on the manifold. Consider the problem of classifying all four-dimensional surfaces in a ten-dimensional ambient space whose metric is spherically symmetric and static. Thus $d=4$, $n=10$, and the symmetry group is $SO(3) \times T^1$ (where translations on $T^1$ correspond to a time shift), so $m=4$. Let us assume that such a metric is by definition representable in the form:
\begin{equation}
	ds^2 = g_{00}(r)\, dt^2 + g_{11}(r)\, dr^2 + r^2 \left( d\tta^2 + \sin^2\tta\, d\vp^2 \right). \label{g_spher}
\end{equation}
We consider both the signature $\left( -,+,+,+ \right)$ and $\left( +,+,+,+ \right)$ as possible ones, because at this level we cannot draw a distinction betwen them. What is essential here is that with such a definition of the symmetry of the metric the action of the group on the manifold is also implicitly defined. To make sure of this, let us use more general definition: a metric is symmetric if the action of the group preserves the distance between any two points of the manifold. The same definition has a coordinate formulation. Let $x^\mu$ be an arbitrarily chosen coordinate system on the manifold and $x'^\mu$ be induced from $x^\mu$ by the action of the element $G$ on the manifold. The metric is symmetric in the sense of distance preservation when $g'_{\mu\nu}$ and $g_{\mu\nu}$ coincide as functions of their argument for each such transformation from $G$. This general definition is equivalent to \eqref{g_spher} in the case where the action of the group $G=SO(3) \times T^1$ on a four-dimensional manifold is defined in the usual way: the coordinates $\tta$, $\vp$ change under the action of $SO(3)$, $t$ changes under the action of $T^1$, and the coordinate $r$ remains unchanged. If the action of the group is specified differently, the equivalence of definitions disappears.

When constructing symmetric surfaces using the method discussed above, the action of the group is determined when we choose which of the coordinates describing the surface should correspond to some group parameters of $G$ and which should not. In order for the induced metric to have the form \eqref{g_spher} and to preserve distances under the surface symmetry transformation (and not under any other), we will restrict the classification problem to four-dimensional surfaces of the second type whose symmetric submanifolds are three-dimensional and numbered with coordinate $r$, so that $\xi=3$. Thereby, we have to deal with the situation $\xi<m$, which restricts the admissible choice of initial vectors to those that remain stationary under the action of some one-parameter stabilizer subgroup. At the same time, since, as mentioned above, the coordinate $t$ corresponds to the group parameter of the subgroup $T^1$ of $G$, the one-parameter stabilizer subgroup of $G$ must be chosen to be a subgroup of $SO(3)$, so it will be isomorphic to $SO(2)$. 

Another advantage of surfaces of the second type is the presence of arbitrariness in the dependence of the initial vector on $r$. In this sense, surfaces of the second type are more diverse than surfaces of the first type, whose appearance is completely determined by the action of the representation on the initial vector of constants. For surfaces of the first type, the induced metric depends in a specific way on $t$ and all three Euler angles parametrizing $SO(3)$. In the case of surfaces of the second type, the metric is surely representable in the form \eqref{g_spher} with functions $g_{00}(r)$ and $g_{11}(r)$, and the symmetry itself does not impose any restrictions on these functions. This allows one to search for different classes for embeddings of any particular metrics -- for example, Schwarzschild or Minkowski spaces.

An analysis of the eigenvalues of the representation matrices shows that an initial vector with a stabilizer subgroup exists for all tensor representations of $SO(3)$ (as part of $G$), but it is absent for all spinor representations with half-integer spin. Therefore, these representations should be excluded from consideration if we are interested in more promising embeddings of the second type. However, as a side note, we consider it appropriate to illustrate the work of the method on the example of the spinor representation of $SO(3)$ with spin $1/2$. For simplicity, we omit the dependence on $t$ for now.

Let $\om\left( \Lambda \right)$ be the mapping $SO(3)\rightarrow SU(2)$ corresponding to such a representation. It is easy to show that the eigenvalues of the matrix $\om$ are $\la = \left(Re\, \om_{11} \right) + i \sqrt{1 - \left( Re\, \om_{11} \right)^2}$ and its complex conjugate $\la^*$. The embedding function of a symmetric surface is obtained by writing in real variables the action $\om\left( \Lambda \right)$ on a two-dimensional complex initial vector $\vec{y}_0$, which we denote as $\vec{y} = \om\left( \Lambda \right)\vec{y}_0$. After realification, we have the following expression for the four-dimensional embedding function:
\begin{equation}
		y^A =
		\begin{pmatrix}
			Re\left( \vec{y} \right)\\
			Im\left( \vec{y} \right)
		\end{pmatrix} =
		\begin{pmatrix}
			Re\left( \om\left( \Lambda \right) \right) & -Im\left( \om\left( \Lambda \right) \right) \\
			Im\left( \om\left( \Lambda \right) \right)  & Re\left( \om\left( \Lambda \right) \right) 
		\end{pmatrix}	
		\begin{pmatrix}
			Re\left( \vec{y}_0 \right)\\
			Im\left( \vec{y}_0 \right)
		\end{pmatrix} \equiv
		V^A{}_B \left( \Lambda \right)\, y_0^B.
\end{equation}
It can be shown that the 4 by 4 matrix $V^A{}_B$ is an element of $SO(4)$ (but not an arbitrary one), and its eigenvalues are the same $\la$ and $\la^*$, the multiplicity of which is increased to two. This means that the stabilizer subgroups equation $V^A{}_B\, y_0^B = y_0^A$ is satisfied only when $Re(\om_{11}) = 1$, which is equivalent to $\om = I$. Hence, there are no vectors in the space of this representation that correspond to a one-parameter stabilizer subgroups, and the surface $y^A$ is three-dimensional rather than two-dimensional, with any choice of the initial vector. But since the action of the $SO(4)$ element preserves the norm of $y_0^A$, locally this surface cannot be anything other than an ordinary three-dimensional sphere in four dimensions. We can introduce angular coordinates on it in which the metric takes the form:
\begin{equation}
		ds^2 = R^2\left( \, d\chi^2 + \sin^2 \chi  \left( d\tta^2 + \sin^2\tta\, d\vp^2 \right) \right), \label{g_3dspher}		
\end{equation}
where the constant $R=y_0^A\, y_0^B\, \de_{AB}$ is the radius of the sphere. As befits a surface of the first type, all three coordinates here have a group nature.

Let us put the coordinate $t$ back in the simplest way, in which $y^a\left( t,\chi,\tta,\vp \right) =\left\{ t,\, y^A\left( \chi,\tta,\vp \right) \right\}$. Such a surface also belongs to the first type, and for it $\xi=d=m=4$. However, if we now substitute $r=R\, \sin \chi$, it turns out that although the surface is not of the second type, the metric nevertheless takes a form consistent with \eqref{g_spher}:
\begin{equation}
		ds^2 = dt^2 + \dfrac{1}{R^2-r^2}dr^2 + r^2 \left( d\tta^2 + \sin^2\tta\, d\vp^2 \right). \label{g_hitroe}		
\end{equation}
As a result, we see that for such a special dependence of radius functions in \eqref{g_spher} the metric (as well as the surface corresponding to it) is symmetric with respect to two differently defined actions of the considered group $G$ on the manifold: both the action which changes all three angles $\chi$, $\tta$, $\vp$, and the other action which leaves the radius $r=R\, \sin \chi$ unchanged. Note that this radius has the same meaning as the radius of the circles $r=R\, \sin \tta$ into which any two-dimensional sphere can be split.

The aim of this paper is to count all symmetric embeddings of the metric \eqref{g_spher} with functions $g_{00}(r)$ and $g_{11}(r)$ whose arbitrariness would not be restricted by the symmetry itself. Therefore, it is further our goal to construct a complete classification of four-dimensional symmetric with respect to $G = SO(3) \times T^1$ surfaces in ten dimensions belonging to the second type, but not to the first. The dimension of symmetric submanifolds $\xi$ here is 3, and the initial vector, as mentioned above, must be chosen in such a way that there exists a stabilizer subgroup $SO(2)$ for it. Let us emphasize once again that such a classification clearly does not include all unsymmetric with respect to $SO(3) \times T^1$ surfaces, the induced metric \eqref{g} on which nevertheless can be symmetric in the sense of representability in the form of \eqref{g_spher}.

\subsection{Method for enumerating surfaces with $SO(3) \times T^1$ symmetry}\label{metod_SO3xT1}
We now proceed directly to the application of the method to the construction of the specified classification. According to it, the embedding function of a symmetric surface can be divided into blocks, each of which is associated with the simplest representations of the group $G$ (usually irreducible) that also correspond to the symmetry group of the ambient space $SO(n_+,n_-)\ltimes T^n$. The basic forms of the blocks corresponding to $G = SO(3) \times T^1$ and $\xi=3$ have already been obtained in the \cite{statja27}, where the case of $n=6$ was solved. Let us briefly describe these results and then apply them to our case $n=10$. To begin with, we assume $r$ and $t$ are fixed and describe the embedding functions symmetric with respect to only $SO(3)$. The discussed requirement of a stabilizer subgroup leads to the fact that for every irreducible tensor representation $SO(3)$ there is a unique (up to a change of basis in the ambient space) realization of the block $y^A$ which is transformed by it. To describe these blocks in a unified way, we introduce a set of bases $\lambda^A_{i_1 ... i_j}$ in tensor spaces of rank $j$, symmetric and traceless over any pair of indices. The irreducible tensor representation $SO(3)$ of spin $j$ and dimension $2j + 1$ will be denoted as $V^j$. Each such representation is associated with a block of the embedding function $y^A(\tta,\vp)$ of the following form:
\begin{align}
	V^0 \quad&\text{---}\quad y^1 = 1, \nonumber\\
	V^1 \quad&\text{---}\quad y^k = \hat{x}^k, \nonumber\\
	V^2 \quad&\text{---}\quad y^A = \lambda^A_{ij}\, \hat{x}^i \hat{x}^j, \nonumber\\
	V^3 \quad&\text{---}\quad y^A = \lambda^A_{ijk}\, \hat{x}^i \hat{x}^j \hat{x}^k, \label{bloki_so3}\\
	\vdots \quad&\nonumber\
\end{align}
where $\hat{x}^i(\tta,\vp) = x^i/r$ is a vector on the unit sphere in spherical coordinates, $A = 1,\ldots,2j + 1$. The choice of the components of the initial vector for each block \eqref{bloki_so3} was dictated by the existence of a stabilizer subgroup. It can be seen that the blocks $y^A$ do correspond to the definition of a symmetric embedding: under the action of a matrix from $SO(3)$ on the vector $\hat{x}^i$, each block $y^A(\tta,\vp)$ in the ambient space is transformed by the corresponding representation of $SO(3)$ (irreducible due to the tracelessness of $\lambda^A_{i_1 . . . i_j}$), and the manifold $y^A$ as a whole transforms into itself. Note also that $y^A(\tta,\vp)$ is nothing but a linear combination of the spherical harmonics $Y_j^m(\tta,\vp)$ of the corresponding spin $j$, and if desired, by choosing the bases $\lambda^A_{i_1 ... i_j}$ one can reduce the components of $y^A$ to the form $y^{j+m+1}=Y_j^m$.

Now let us include the symmetry of $T^1$ with respect to time translations. Any suitable representation of the group $SO(3) \times T^1$ with elements $g = \left\{O, t\right\}$ can be represented as a direct product of the following form:
\begin{equation}
	V(g) = V^j(O) \otimes S^s (\gm t) \otimes P^p (\al t) \otimes Q^q (\bt t), \label{predst}
\end{equation}
where $\gm, \al, \bt$ are arbitrary dimensional parameters; $j = 0, 1,\ldots$ -- the spin of the $SO(3)$ representation; parameter $s = 0, 1,\ldots$; binary parameters $p, q = 0,1$ are responsible for the inclusion of the corresponding multipliers; quantities with zero indices $S^0 = P^0 = Q^0 = 1$, and otherwise have the form of matrices:

\begin{equation}
	S^1(\gm t) =
	\begin{pmatrix}
		1 & \gm t\\
		0 & 1
	\end{pmatrix}, \quad
	S^2(\gm t) =
	\begin{pmatrix}
		1 & \gm t & \frac{\left( \gm t \right)^2}{2!}\\
		0 & 1 & \gm t\\
		0 & 0 & 1
	\end{pmatrix}, \quad
	S^3(\gm t) =
	\begin{pmatrix}
		1 & \gm t & \frac{\left( \gm t \right)^2}{2!} & \frac{\left( \gm t \right)^3}{3!}\\[2pt]
		0 & 1 & \gm t & \frac{\left( \gm t \right)^2}{2!}\\[2pt]
		0 & 0 & 1 & \gm t\\[2pt]
		0 & 0 & 0 & 1
	\end{pmatrix}, \quad \dots
	\label{S_matrices}
\end{equation}

\begin{equation}
	P^1 (\al t) = 
	\begin{pmatrix}
		\cos \al t & -\sin\al t \\
		\sin \al t & \cos \al t \\
	\end{pmatrix}, \quad 	
	Q^1 (\bt t) = 
	\begin{pmatrix}
		\cosh \bt t & \sinh\bt t \\
		\sinh \bt t & \cosh \bt t \\
	\end{pmatrix}.
\end{equation}
We will not describe in detail the derivation of \eqref{predst}, but its general logic is as follows: the representation of $T^1$ is written as an exponent from an arbitrary linearly $t$-dependent matrix, which can be reduced to the form of a single Jordan block. The exponent from a matrix with units above the diagonal reduces to $S^s$, and the exponent from a unit matrix with arbitrary complex multiplier $\left( \bt + i \al \right)t$ generates matrices $P^1$ (realification of the complex part) and $Q^1$ (the form of the real part satisfying the problem). Note that $Q^1$ is the simplest suitable representation, but it is reducible, unlike all the others.

Embeddings with the required symmetry consist of blocks, each of which is obtained as a result of the action of one of the representations \eqref{predst} on the initial vectors. In the most general form, the components of these blocks can be written as:
\begin{equation}
	y^A_{FMN}\left( t, r, \tta, \vp \right) = y^A\left(\tta, \vp \right) S^s (\gm t)_{F F'}\, P^p (\al t)_{M M'}\, Q^q (\bt t)_{N N'}\, C_{F'M'N'}(r),
\end{equation}
where $F, F' = 1, \ldots, s+1$; $M, M', N, N' = 1,2$; $C_{F'M'N'}(r)$ is a set of arbitrary functions of radius (this is a part of the initial vector; the other part is absorbed into $y^A\left(\tta, \vp \right)$ according to \eqref{bloki_so3}). Each of these blocks is parameterized by four integers:
\begin{equation}
	\left\{j = 0, 1, 2, \ldots\,|\, s = 0, 1, 2, \ldots\,|\, p = 0,1\,|\, q = 0,1 \right\},
\end{equation}
which are the classifiers of all possible blocks that can be used to form the embeddings of a given symmetry. In addition, the embeddings can differ by three dimensional parameters $\al, \bt, \gm$ and some number of arbitrarily defined radius functions.

In the case of $\left\{ j, 0, 0, 0 \right\} $ we have blocks $y^A(r, \tta, \vp)$ transformed by pure tensor representations $SO(3)$, i.e. having the form \eqref{bloki_so3} with additional radius-dependent coefficients. To show what the numbers $s,p,q$ are responsible for, let us write out the three simplest blocks corresponding to them:

\begin{align}
	\left\{ 0, 1, 0, 0 \right\} \quad&\text{---}\quad y^0 = f(r) + \gm t, 
	\label{blok_t}\\
	\left\{ 0, 0, 1, 0 \right\} \quad&\text{---}\quad y^{0,1} = f(r) 
	\begin{pmatrix}
		\sin\left( \al t + \psi(r) \right)\\
		\cos\left( \al t + \psi(r) \right)
	\end{pmatrix}, 
	\label{blok_sin}\\
	\left\{ 0, 0, 0, 1 \right\} \quad&\text{---}\quad y^{0,1} = f(r) 
	\begin{pmatrix}
		\sh\left( \bt t + \psi(r) \right)\\
		\ch\left( \bt t + \psi(r) \right)
	\end{pmatrix}. 
	\label{blok_sh}
\end{align}
Here and hereinafter $f(r)$, $\psi(r)$, $\gm$, $\al$, and $\bt$ are notations for arbitrary functions of radius and constants, not necessarily the same for different blocks. The components of the blocks $\left\{ 0, s, 0, 0 \right\}$ with $s>1$ are polynomials in $t$ of degree $s$.

We will call the mixing of the tensor representation with time the case when both $j$ and at least one of the numbers $s,p,q$ are different from zero. For example, the six-dimensional block $\left\{ 1, 0, 1, 0 \right\}$ can be represented in the form analogous to \eqref{blok_sin} as a direct product by substituting $f(r)\, \hat{x}^i(\tta, \vp)$ in place of $f(r)$. Mixing is possible with nonzero $p, q$, nonzero even $s$, and with any combination of them. The case of odd $s$ is exceptional, because in the full direct sum of representations only one term with odd $s$ is allowed, and for it $j, p, q$ must be zero, so mixing is forbidden. This is due to the fact that for odd $s$ the matrices \eqref{S_matrices} cannot realize only rotations in the ambient space, but necessarily generate translations as well.

Note that the embeddings that do not contain any blocks with odd spin $j$ have an additional symmetry with respect to the replacement of $x^i$ by $-x^i$, and hence their sections $\left\{t, r = Const\right\}$ do not have the topology of a sphere. Specific physical problems, such as, for example, the searching for exact explicit embeddings of spherically symmetric black holes, can impose restrictions on the topology, and the corresponding embeddings in them should be excluded from consideration. However, we will not avoid such embeddings, keeping in mind that even if the physical problem admits only the topology of a sphere, these embeddings can be used as a background for some perturbation theory. Then the corrections can save the embedding from such unwanted symmetry.

Finally, we introduce dimension-related auxiliary notation for blocks of the embedding function in addition to the somewhat cumbersome notation $\left\{ j, s, p, q \right\}$. Blocks that transform by purely tensor (without mixing) representations, we will denote simply by the number $m=2j+1$ equal to their dimension. The blocks $\left\{ 0, s, p, q \right\}$, i.e. scalars of $SO(3)$, we divide by parity of the number $s$ into two sets of groups. For even $s$, the dimension of such blocks is $l = (s+1)\, 2\,{}^{p+q}$, and we will denote them as $ \left< l \right>$ (we emphasize that this is actually a group of different blocks of the same dimension; in the old notation they are still distinguishable). In the case of odd $s$, the block dimension $l = s$, and the notation will be $\{ l \}$. The case of mixing tensor and time representations will be denoted as $m \times \left< l \right>$ (as mentioned above, the case $m \times \{ l \}$ is forbidden).

In this way, all possible blocks are described, and any embedding with $SO(3) \times T^1$ symmetry must be represented in terms of a direct sum of them.
\section{Classification and properties of ten-dimensional symmetric embeddings} \label{rabota}
Using the results given in the previous Section, we can classify all embeddings of the metric \eqref{g_spher} in the ten-dimensional $R\,{}^{n_+,n_-}$ ambient space possessing $SO(3) \times T^1$ symmetry not only with respect to the metric, but also with respect to the four-dimensional surface itself. All such embedding functions are a direct sum of blocks with and without mixing, such that their total dimension is 10. Given that the simplest block with mixing $3\times \left<2\right>$ already has dimension 6, two blocks with mixing cannot fit into the required dimension. Therefore, all possible embeddings can be represented as $\{ l \} + m_0\times\left<l_0\right>+ \left<l_1\right> + \ldots + m_1 + \ldots$, where the numerical value of this sum is 10 when brackets are omitted, and the simultaneous presence of terms of all types is not necessary. We will not give a complete classification for arbitrary $n_+$ and $n_-$, but all 52 cases corresponding to the $(1,9)$ signature will be summarized in a single table at the end of this Section. Before that, let us discuss two properties important in the context of the problems in the Regge-Teitelboim embedding theory that the corresponding classes may or may not possess. These are the admissibility of flat embeddings and the unfolding property already mentioned in the introduction.
	
By a flat embedding we mean an embedding of the Minkowski space for which $g_{00}(r)=-1$ and $g_{11}(r)=1$ in the formula \eqref{g_spher}. Flat symmetric embeddings are useful as a background in the context of various perturbation theory problems arising in embedding theory. Checking whether a particular class contains such an embedding is straightforward: the functions $g_{00}$ and $g_{11}$ are expressed in terms of the functions of radius arising from the construction of $y^a$, and then looking at the equations themselves one can see whether it is possible to satisfy them at least locally by fixing arbitrariness. At the same time, as a rule, such a flat background is also required to be unfolded and smooth at all values of coordinates. We will discuss unfolding and methods of its verification further on. As for smoothness, proving the existence of smooth solutions, and especially finding their explicit form, requires additional analysis, which is beyond the scope of our paper. Therefore, we restrict ourselves to denote in the final table for the signature $\left(1,9\right)$ all those classes for which it is possible to prove in some way the absence of smooth unfolded Minkowski embeddings. We will not dwell on it in detail, but as an illustration let us mention the class $\{1\} + \left<2\right> + 3 \times \left<2\right> + 1$. It is easy to show that flat embeddings of this class exist only inside a ball of finite radius (outside it $g_{00}(r)$ grows indefinitely), hence there is no smoothness for all coordinates. In the final table, we will note all those 8 classes that we did not reject according to any of the listed criteria.
	
Let us move on to the unfolding, which requires a more detailed discussion. Its definition is related to the properties of the second fundamental form of the surface $b^a_{\mu\nu}$:
\begin{equation}
	b^a_{\mu\nu} = D_\mu \partial_\nu y^a = \partial_\mu \partial_\nu y^b\,  \Pi_{\bot}{}^a_b.
	\label{b}
\end{equation}
Here $D_\mu$ is the covariant derivative consistent with the metric, $\Pi_{\bot}{}^a_b$ is the projector onto a subspace orthogonal to the surface. Since the lower indices of $b^a_{\mu\nu}$ are symmetric, they can be replaced by a single multiindex that runs over 10 values. Taking into account the projection onto an orthogonal subspace, one can see that the index $a$ effectively runs through 6 values. So $b^a_{\mu\nu}$ is representable in the form of some $10\times6$ matrix. An embedding is called unfolded if this matrix has rank 6 which is a maximal possible rank of it.

Let the multiindex $A = \left\{\mu, \nu\right\}$ number the rows of the matrix, the index $a$ number its six columns, i.e., $b_{Aa}$ corresponds to the components of $b^a_{\mu\nu}$ in a fixed basis (in which the components $a=6,7,\ldots$ are zero). Transition to another coordinate system for $b_{Aa}$ means multiplication from the left by the reversible $10\times10$ transition matrix. The rank cannot change, and hence the definition of unfolding remains coordinate invariant. By construction, the embeddings symmetric with respect to $SO(3) \times T^1$ are written in spherical coordinates (at least up to substitution $r'=f(r)$). In the same coordinates it is most convenient to check the rank of $b_{Aa}$. 

Direct proof of the presence of unfolding involves a cumbersome procedure of calculating the rank of $10\times6$ matrices, so this work was done using computer algebra systems. However, a much simpler task is to prove the absence of unfolding rather than its presence. Below we first list several classes for which we can show analytically that the rank of $b_{Aa}$ is known to be less than 6, i.e. there is no unfolding. Then, in the last paragraph, we describe the method of numerical computation of the rank and give its results for all classes of signature $(1,9)$, including those for which there is no analytic proof.

\subsection{Class  $\{l\} + \left<l_1\right> + \ldots + \left<l_N\right> + 1 + \ldots + 1 + 3$}
The embedding function of this class has the form:
\begin{equation}
	y^a = 
	\begin{pmatrix}
		y^0(r,t)\\
		\vdots\\
		h_1(r)\\
		\vdots\\
		f(r) \hat{x}^k(\tta,\vp)
	\end{pmatrix}.
\end{equation}
Embeddings with only one $\{7\}$ block or only $\left<l_i\right>$ blocks in addition to block $3$ also belong to this class. Suppose, without the loss of generality, that in a subspace orthogonal to the surface we can introduce a basis $V^a_N$, $N = 1, \ldots, 6$, orthonormalized on $\pm 1$. The projector is then represented as:
\begin{equation}
	\Pi_{\bot}{}^{ab} = V^a_M\, V^b_N\, \eta^{MN},
\end{equation}
where $\eta^{MN}$ is a metric with $\pm 1$ on the diagonal, corresponding to the subspace signature. For convenience, we introduce a uniform notation for the angular variables: $\vp_j \in \left\{\tta, \vp \right\}$, $j=1,2$. From the condition of orthogonality of the basis to the surface, we obtain:
\begin{equation}
	V_N^a\, \partial_{\vp_j} y_a = f(r)\, V^{6+k}_N \, \partial_{\vp_j} \hat{x}_k = 0, \quad j=1,2.
\end{equation}
Six three-dimensional vectors $V^{6+k}_N$ turn out to be orthogonal to two vectors tangent to the sphere in the corresponding three-dimensional subspace. This means that they are all radial, that is, $V^{6+k}_N = C_N\, \hat{x}^k$. As can be seen, the indices N and k factorize in some sense.

Consider now 7 of 10 rows of the matrix $b_{Aa}$ with indices $A = \left\{ \mu, \vp_j \right\}$:
\begin{equation}
	b^a_{\mu\vp_j} = V^a_M\, V^{6+k}_N\, \eta^{MN} \partial_\mu \partial_{\vp_j} y_{6+k}= \left( \hat{x}^k  \partial_\mu \partial_{\vp_j} y_{6+k}  \right) C_N \eta^{NM} V^a_M , \quad j = 1, 2.
\end{equation}
These seven rows are found to be proportional to one row vector $C_N \eta^{NM} V^a_M$, hence only four rows of $b_{Aa}$ out of ten can be chosen as linearly independent. In other words, the matrix $b_{Aa}$ has a rank no higher than four, so there is no unfolding in this class.

\subsection{Class $\{1\} + m + 1 + \ldots + 1$,\; $m = 3, 5, 7, 9$} \label{1+m+1+...}
This is a class of embeddings with one component that is linear w.r.t. time, one tensor block, and the number of scalar components necessary to achieve the desired dimension:
\begin{equation}
	y^a = 
	\begin{pmatrix}
		t\\
		f(r)\, y^A(\tta,\vp)\\
		h_1(r)\\
		\vdots
	\end{pmatrix},
\end{equation}
where $y^A(\tta,\vp)$ corresponds to \eqref{bloki_so3}. The embedding without scalars $\{1\} + 9$ also belong to this class. 

The absence of unfolding here is due to the factorization of the angular and radial variables when taking the corresponding second derivatives. Because of this, two lines $b_{Aa}$ at $A = \left\{ r, \vp_j \right\}$ are zero:
\begin{equation}
	b^a_{r\vp_j} = \Pi_{\bot}{}^a_A\, f'(r)\, \partial_{\vp_j} y^A(\tta,\vp) = \frac{f'}{f}\, \Pi_{\bot}{}^a_b\, \partial_{\vp_j} y^b = 0, \quad j = 1, 2.
\end{equation}
In addition to these two lines, the four lines $A = \left\{ t, \mu \right\}$ are automatically zero. Thus, the rank of $b_{Aa}$ is at most four, and there is no unfolding. 

\subsection{Class $\left<2\right> + m + 1 + \ldots + 1$,\; $m = 3, 5, 7$} \label{2+m+1+...}
The case differs from the previous one in the fact that the single t-dependent component is replaced by a two-dimensional block with ordinary or hyperbolic sines and cosines (for certainty, will consider the ordinary ones):
\begin{equation}
	y^a = 
	\begin{pmatrix}
		g(r) \sin\left( \al t + \psi(r) \right)\\		
		g(r) \cos\left( \al t + \psi(r) \right)\\
		f(r)\, y^A(\tta,\vp)\\
		h_1(r)\\
		\vdots
	\end{pmatrix}.
\end{equation}

For the same reasons as in the previous case, we can say that the lines $r\tta$ and $ r\vp$, as well as the lines $t\tta$ and $t\vp$, are zero. Note now that the three derivatives $\partial_t y^a$, $\partial_t \partial_t y^a$, and $\partial_t \partial_r y^a$ are in the same two-dimensional subspace, and hence linearly dependent. Consequently, after projecting to a direction orthogonal to $\partial_t y^a$, the rows $tt$ and $tr$ will also be linearly dependent, which reduces the rank of $b_{Aa}$ to five and leads to the absence of unfolding.

\subsection{Classes $\{1\} + 3 + 3 + 1 + 1 + 1$ and $\{1\} + 3 + 3 + 3 $} \label{1+3+3+...}
To be specific, consider an embedding of the first class (there are no differences in logic for the second):
\begin{equation}
	y^a = 
	\begin{pmatrix}
		t\\
		f_1(r) \hat{x}^k(\tta,\vp)\\
		f_2(r) \hat{x}^k(\tta,\vp)\\
		h_1(r)\\
		\vdots
	\end{pmatrix}.
\end{equation}

The four lines of $b_{Aa}$ corresponding to $t\mu$ are zero together with the corresponding second derivatives. In addition to them, as in the cases \ref{1+m+1+...} and \ref{2+m+1+...}, some factorization takes place, resulting in a fifth zero line. However, the factorization is a little more complicated here. To see it, we fix the definition of angles as follows:
\begin{equation}
	\hat{x}^k = 
	\begin{pmatrix}
		\sin\tta \cos\vp\\
		\sin\tta \sin\vp\\
		\cos\tta
	\end{pmatrix}.
\end{equation}
Then it is easy to show by straightforward calculation that when taking the derivative of $\vp$ from the embedding function, the factorization of the variables appears in the following form:
\begin{equation}
	\partial_\vp y^a \equiv \sin \tta\, X^a(r,\vp),
\end{equation}
so the string $A = \left\{\tta, \vp \right\}$ turns out to be zero:
\begin{equation}
	b^a_{\tta\vp} = \frac{\cos \tta}{\sin \tta}\, \Pi_{\bot}{}^a_b\, \partial_\vp y^b = 0,
\end{equation}
therefore, the rank of $b_{Aa}$ is obviously no higher than five, and there is no unfolding in these two classes.

\subsection{Unfolding of background embeddings of the nonrelativistic limit in the embedding theory}\label{1+m}
When analyzing the equations of the embedding theory, a particular class of embeddings in an ambient space with signature $(1,9)$ is of special interest, namely, embeddings of the form $\{ 1 \} + m_1 + \ldots + m_N$. For them, the time $t$ enters only one component of the embedding function, and linearly. It is the embeddings possessing such a property that are used as a background both in the paper \cite{Kuptsov_2022} and in the study of the nonrelativistic limit of the embedding theory in the paper \cite{statja68}. In total there are 7 types of such embeddings:
\begin{align}
	\{ 1 \} &+ 3 + 1 + 1 + 1 + 1 + 1 + 1,\label{1}\\
	\{ 1 \} &+ 3 + 3 + 1 + 1 + 1,\label{2}\\
	\{ 1 \} &+ 3 + 3 + 3,\label{3}\\
	\{ 1 \} &+ 5 + 1 + 1 + 1 + 1,\label{4}\\
	\{ 1 \} &+ 5 + 3 + 1,\label{razvernut}\\
	\{ 1 \} &+ 7 + 1 + 1,\label{6}\\
	\{ 1 \} &+ 9.\label{7}
\end{align}

For them, the nontrivial part of the second fundamental form $b^a_{\mu\nu}$ is a symmetric on three-dimensional indices quantity $b^a_{i j}$. By analogy with what was said earlier, it can be represented as a square $6\times6$ matrix, and the unfolding will then be equivalent to the nondegeneracy of this matrix. In \cite{Kuptsov_2022} it was shown that for an embedding $\{1\}+5+3+1$ the quantity $b^a_{i j}$, understood as a $6\times6$ matrix, is in general reversible, and hence the corresponding embedding is unfolded. For the remaining six variants from \eqref{1}-\eqref{7} there is no unfolding, which follows from \eqref{1+m+1+...} and \eqref{1+3+3+...}. Thus, the embedding $\{1\}+5+3+1$ is distinguished from the embeddings where time enters only linearly as the only unfolded one.

\subsection{Full classification in the case of signature $(1,9)$}
In the table \ref{table} we list all possible 52 cases of symmetric with respect to $SO(3) \times T^1$ four-dimensional embeddings in $R\,{}^{1,9}$ whose induced metric is represented in the form \eqref{g_spher} with arbitrary functions $g_{00}(r)$ and $g_{11}(r)$. For each of them the rank of the matrix $b_{Aa}$ constructed by the second fundamental form is given, which was computed numerically in the general situation (about the technique see below). Only embeddings from classes with rank 6 are unfolded in the general case. All classes were checked for admissibility of everywhere smooth unfolded Minkowski embeddings. The 8 classes that pass all our checks are marked in gray. All other classes cannot contain such embeddings. The construction of the explicit form of such embeddings in these eight classes, as mentioned above, is beyond the scope of this paper.

Let us clarify the notation used in the table. When describing the blocks of an embedding function from a class written in the notations $\{l\}, \left<l\right>, m$, an ambiguity may arise, which was absent in the more informative but also more cumbersome notations $\left\{ j, s, p, q \right\}$ (for more details about the notations, see the end of the Section \ref{metod_SO3xT1}). To achieve certainty, let us fix the order of the signature signs as $\left( -,+,\ldots,+ \right)$ and agree to distinguish classes by the order of the terms. For the $(1,9)$ signature (but not for all others), this is enough to remove the ambiguity. So, for example, the components of the $\{1\}+\left<2\right>+\ldots$ embedding include $t$, $\sin$, and $\cos$, but the $\left<2\right>+\{1\}+\ldots$ embedding includes $\sh$, $\ch$, and $t$ (these are analogs of the Davidson-Paz \cite{davidson} embedding).

\begin{table}[t]
	\centering
	\begin{tabular}{ | l | c| }
		\hline
		\multicolumn{1}{|c|}{Embedding class} & Rank\\
		\hline
		$\{1\} + 3 + 1 + 1 + 1 + 1 + 1 + 1 $&2 \\
		$\{1\} + 3 + 3 + 1 + 1 + 1 $&4\\
		$\{1\} + 3 + 3 + 3 $&4\\
		$\{1\} + 5 + 1 + 1 + 1 + 1 $&4\\
		\rowcolor{light-gray}
		$\{1\} + 5 + 3 + 1$&6\\
		$\{1\} + 7 + 1 + 1 $&4\\
		$\{1\} + 9$ &2\\
		$\{1\} + \left<2\right> + 3 + 1 + 1 + 1 + 1$ &4\\
		$\{1\} + \left<2\right> + 3 + 3 + 1$&6\\
		$\{1\} + \left<2\right> + 5 + 1 + 1$&6\\
		$\{1\} + \left<2\right> + 7$&4\\
		$\{1\} + \left<2\right> + \left<2\right> + 3 + 1 + 1$&4\\
		\rowcolor{light-gray}
		$\{1\} + \left<2\right> + \left<2\right> + 5$&6\\
		$\{1\} + \left<2\right> + \left<2\right> + \left<2\right> + 3$&4\\
		$\{1\} + 3 \times \left<2\right> + 1 + 1 + 1$&5\\
		$\{1\} + 3 \times \left<2\right> + 3$&6\\
		$\{1\} + 3 \times \left<2\right> + \left<2\right> + 1$&6\\
		$\left<2\right> + 3 + 1 + 1 + 1 + 1 + 1$&3\\
		$\left<2\right> + 3 + 3 + 1 + 1$&5\\
		$\left<2\right> + 5 + 1 + 1 + 1$&5\\		
		\rowcolor{light-gray}
		$\left<2\right> + 5 + 3$&6\\
		$\left<2\right> + 7 + 1$&4\\
		$\left<2\right> + \left<2\right> + 3 + 1 + 1 + 1$&4\\
		$\left<2\right> + \left<2\right> + 3 + 3$&6\\
		\rowcolor{light-gray}
		$\left<2\right> + \left<2\right> + 5 + 1$&6\\
		$\left<2\right> + \left<2\right> + \left<2\right> + 3 + 1$&4\\
		\hline
	\end{tabular}
	\;
	\begin{tabular}{ | l | c| }
		\hline
		\multicolumn{1}{|c|}{Embedding class}  & Rank \\
		\hline
		$\left<2\right> + 3 \times \left<2\right> + 1 + 1$&6\\
		\rowcolor{light-gray}
		$\left<2\right> + 3 \times \left<2\right> + \left<2\right>$&6\\
		$\left<2\right> + \{1\} + 3 + 1 + 1 + 1 + 1$&4\\
		$\left<2\right> + \{1\} + 3 + 3 + 1$&6\\
		\rowcolor{light-gray}
		$\left<2\right> + \{1\} + 5 + 1 + 1$&6\\
		$\left<2\right> + \{1\} + 7$&4\\
		$\left<2\right> + \{1\} + \left<2\right> + 3 + 1 + 1$&4\\
		\rowcolor{light-gray}
		$\left<2\right> + \{1\} + \left<2\right> + 5$&6\\
		$\left<2\right> + \{1\} + \left<2\right> + \left<2\right> + 3$&4\\
		$\left<2\right> + \{1\} + 3 \times \left<2\right> + 1$&6\\
		$\left<3\right> + 3 + 1 + 1 + 1 + 1$&4\\
		$\left<3\right> + 3 + 3 + 1$&6\\
		$\left<3\right> + 5 + 1 + 1$&6\\
		$\left<3\right> + 7$&4\\
		$\left<3\right> + \left<2\right> + 3 + 1 + 1$&4\\
		$\left<3\right> + \left<2\right> + 5$&6\\
		$\left<3\right> + \left<2\right> + \left<2\right> + 3$&4\\
		$\left<3\right> + 3 \times \left<2\right> + 1$&6\\
		$\{3\} + 3 + 1 + 1 + 1 + 1$&4\\
		$\{3\} + 3 + 3 + 1$&6\\
		$\{3\} + 5 + 1 + 1$&6\\
		$\{3\} + 7$&4\\
		$\{3\} + \left<2\right> + 3 + 1 + 1$&4\\
		\rowcolor{light-gray}
		$\{3\} + \left<2\right> + 5$&6\\
		$\{3\} + \left<2\right> + \left<2\right> + 3$&4\\
		$\{3\} + 3 \times \left<2\right> + 1$&6\\
		\hline
	\end{tabular}
	\caption{Classification of all symmetric embeddings in $R\,{}^{1,9}$ and their corresponding rank of the matrix $b_{Aa}$; gray marks all those classes which can contain the unfolded and everywhere smooth embeddings of the Minkowski metric.}
	\label{table}
\end{table}

Some classes given in the table correspond to the cases described earlier, i.e. at least the upper bound of the rank for them is defined analytically. However, for these cases, as well as for all others, the exact calculation of the rank by analytical methods is unreasonable due to the cumbersomeness of the procedure, so the analysis was carried out using computer algebra systems. The method of calculation was as follows: first, the embedding function $y^a$ corresponding to each class was defined, including a number of arbitrary functions of radius; the basis of tensors $\lambda^A_{i_1 ... i_j}$ was chosen for convenience in such a way that the tensor blocks $y^A$ represent spherical harmonics\texttt{} (see \eqref{bloki_so3} and the remark right after); the second fundamental form $b^a_{\mu\nu}$ was calculated from $y^a$, and the matrix $b_{Aa}$ was formed from its components; finally, all coordinates and radius functions were interpreted as taken at a random numerical point, i.e. all functions and their derivatives were replaced by independently randomized numbers, and the rank was calculated for the already numerical matrix $b_{Aa}$. The value was considered reliable after multiple repetitions of the procedure (for cases where the embedding function included $\sh$ and $\ch$, the algorithm could sometimes underestimate the rank, which we explain by its incorrect work with exponentially large numbers; the problem was solved by narrowing closer to zero the range of randomization of expressions under these functions).

Finally, we note that 4 groups of 8 classes can be identified from the table, which differ only in the initial three components: $\{1\}+\left<2\right>$, $\left<2\right>+\{1\}$, $\left<3\right>$, and $\{3\}$. The ranks in them turned out to depend only on the remaining terms, so the sequence of ranks in the table is repeated for each of these groups.

\section{Conclusion}
Using the method proposed in the \cite{statja27} for constructing embeddings with a given symmetry, we have shown that any $SO(3)\times T^1$-symmetric embedding of four-dimensional space-time with a metric of the form \eqref{g_spher} into a ten-dimensional pseudo-Euclidean space can be enumerated by a string composed of natural numbers of the form $\{ l \} + m_0\times\left<l_0\right>+ \left<l_1\right> + \ldots + m_1 + \ldots$, which reflects the structure of the corresponding group representation. Each term in the line corresponds to a block of the embedding function of the same dimension: the block of odd dimension $\{ l \}$ consists of polynomials on $t$ of degree $l$, some of whose coefficients depend on $r$ in an arbitrary way; the block of even dimension $\left<l\right>$ consists of either similar polynomials, or pairs of sine and cosine (ordinary/hyperbolic) functions on $t$ and $r$, or a combination of both; the block $m=2j+1$ consists of linear combinations of spherical harmonics of degree $j$; the blocks $m\times\left<l\right>$ are obtained as a tensor product of those described above; in addition, the length of each block as a whole may depend arbitrarily on $r$. The case of the ambient space of signature $(1,9)$ admits 52 classes of symmetric embeddings, and all of them are given in the table \ref{table}.

A significant part of the paper has been devoted to the analysis of unfolding \cite{Zaitseva_2021} of the embeddings for different classes, which is motivated by the importance of this property in solving actual problems of the Regge-Teitelboim embedding theory, such as the study of properties of fictitious embedding matter in the nonrelativistic regime of its motion \cite{statja68}. If the embedding is unfolded, it means that the second fundamental form \eqref{b} in some sense is non-degenerate, which can be used to simplify the equations of motion. For some classes of symmetric embeddings it was possible to prove analytically the absence of unfolding.
In particular, among several classes of type $\{ 1 \} + m_1 + \ldots + m_N$ that are significant in the context of the nonrelativistic limit of the embedding theory, only the embeddings of the type $\{ 1 \} + 5 + 3 + 1$ are unfolded. Such uniqueness is an additional justification for choosing this particular type of embedding as a background for the linearized Regge-Teitelboim equations in \cite{Kuptsov_2022}.

The choice of embeddings $\{ 1 \} + m_1 + \ldots + m_N$ as a background leads to the regime of motion, which in \cite{statja68} was named as "non-relativistic in the bulk"\ limit. This, however, is only a sufficient, but not necessary condition that the embedding matter (understood as a part of degrees of freedom of the embedding function) moves in a nonrelativistic manner. Therefore, though the question of choosing a symmetric background is closed for nonrelativism in the bulk, at more general analysis of nonrelativistic embedding matter other symmetric embeddings of signature $(1,9)$ can be useful. The check of the unfolding property for them has been carried out with application of computer algebra systems, and it has shown that only 23 classes out of 52 allow unfolding. In addition, the classes were checked for the presence of flat and smooth at each point embeddings, which also plays a role in the choice of background. As a result, we found 8 classes that could admit simultaneously flat, unfolded, and everywhere smooth symmetric embeddings. These classes are marked in gray in the table \ref{table}. 

{\bf Acknowledgements.}
The work of S.~S.~Kuptsov was supported by the Ministry of Science and Higher Education of
the Russian Federation, agreement no. 075-15-2022-289.

\end{document}